\documentclass[reprint,groupedaddress,amsmath,amssymb,aps,pra,floatfix]{revtex4-1}
\pdfoutput=1
\usepackage{amsmath,amssymb,graphicx,mathrsfs,amsfonts}

\newcommand{\ket}[1]{\ensuremath{\left|  #1 \right\rangle}}

\draft % marks overfull lines with a black rule on the right

\begin{document}

%\title{Nondestructive Strontium Counting with a Strongly Coupled Cavity QED System on a Forbidden Transition} %Title of paper
\title{Narrow-line Laser Cooling by Adiabatic Transfer} %Title of paper



\author{Matthew A. Norcia}
\affiliation{JILA, NIST, and University of Colorado, 440 UCB, 
Boulder, CO  80309, USA}
\author{Julia R. K. Cline}
\affiliation{JILA, NIST, and University of Colorado, 440 UCB, 
Boulder, CO  80309, USA}
\author{John P. Bartolotta}
\affiliation{JILA, NIST, and University of Colorado, 440 UCB, 
Boulder, CO  80309, USA}
\author{Murray J. Holland}
\affiliation{JILA, NIST, and University of Colorado, 440 UCB, 
Boulder, CO  80309, USA}
\author{James K. Thompson}
\affiliation{JILA, NIST, and University of Colorado, 440 UCB, 
Boulder, CO  80309, USA}
\email[]{matthew.norcia@colorado.edu}
%\thanks{}
%\altaffiliation{NIST,325 Broadway, Boulder, CO 80305, USA}
%\affiliation{JILA, NIST, and University of Colorado, 440 UCB, 
%Boulder, CO  80309, USA}

\date{\today}

\begin{abstract}
We propose and demonstrate a novel laser cooling mechanism applicable to particles with narrow-linewidth optical transitions.  By sweeping the frequency of counter-propagating laser beams in a sawtooth manner, we cause adiabatic transfer back and forth between the ground state and a long-lived optically excited state.  The time-ordering of these adiabatic transfers is determined by Doppler shifts, which ensures that the associated photon recoils are in the opposite direction to the particle's motion. This ultimately leads to a robust cooling mechanism capable of exerting large forces via a weak transition and with reduced reliance on spontaneous emission.  We present a simple intuitive model for the resulting frictional force, and directly demonstrate its efficacy for increasing the total phase-space density of an atomic ensemble.  We rely on both simulation and experimental studies using the 7.5~kHz linewidth $^1$S$_0$ to $^3$P$_1$ transition in $^{88}$Sr.  The reduced reliance on spontaneous emission may allow this adiabatic sweep method to be a useful tool for cooling particles that lack closed cycling transitions, such as molecules.
\end{abstract}


\maketitle 

%%%%%%%%%%%%%%%%% Intro %%%%%%%%%%%%%
The development of Doppler cooling and sub-Doppler cooling in the 1980's has revolutionized our ability to control neutral atoms, ions, and mechanical resonators \cite{chu1986experimental, diedrich1989laser, chan2011laser}.  
%Since then, new forms of laser cooling have enabled its extension into new parameter regimes \cite{PhysRevLett.82.1116, kerman2000beyond}, and to new types of systems \cite{chan2011laser, shuman2010laser}.  
Since then, Doppler cooling techniques have been extended to narrow-linewidth optical transitions to achieve lower temperatures and high phase-space density \cite{PhysRevLett.82.1116, vogel1999narrow, PhysRevLett.110.263003, PhysRevA.84.061406, PhysRevA.61.061403, PhysRevLett.93.073003}. Additionally, there is an ongoing effort to use Doppler cooling for molecules \cite{shuman2010laser, hummon20132d}.  Each of these pursuits face certain persistent limitations.  The use of weak transitions limits the forces achievable with Doppler cooling, and the narrow linewidth of the transition makes the cooling very sensitive to perturbations of the cooling transition frequency \cite{doi:10.1143/JPSJ.68.2479} and technical drifts in the cooling laser frequency.  In the case of molecules, the large number of spontaneous emissions required for Doppler cooling is a key obstacle to creating ensembles of ultra-cold molecules due to the high probability of spontaneous Raman transitions to undesired internal states.

Here we present a new form of laser cooling that mitigates these issues.  Our technique relies on the adiabatic transfer of atoms to and from a long-lived optically excited state to both slow and cool the atoms.  Because the role of spontaneous emission is reduced (though not eliminated) relative to standard Doppler cooling techniques, our technique enables large forces to be generated even on weak transitions, and may facilitate the extension of laser cooling techniques to systems that lack closed cycling transitions.  Of particular interest, this technique may be applied to the slowing and cooling of molecules with narrow linewidth optical transitions \cite{collopy2015prospects, kobayashi2014prospects}.  

From a more fundamental perspective, this work has important implications for the ongoing discussion of the role of spontaneous emission in dissipating entropy during laser cooling \cite{metcalf2008entropy, corder2015laser}.  Our view is that spontaneous emission (or another form of dissipation) is necessary to achieve phase-space compression, but that the total number of spontaneously scattered photons required to do so can be quite low (of order one).  Similar conclusions have been reached in the study of optical pumping in high angular momentum states of atoms \cite{PhysRevA.94.043416} and cooling with single spontaneous emission events \cite{price2008single}.

%%%%%%%%%%%%%%%%% Fig 1 %%%%%%%%%%%%%

\begin{figure}[!htb]
\includegraphics[width=3.5in]{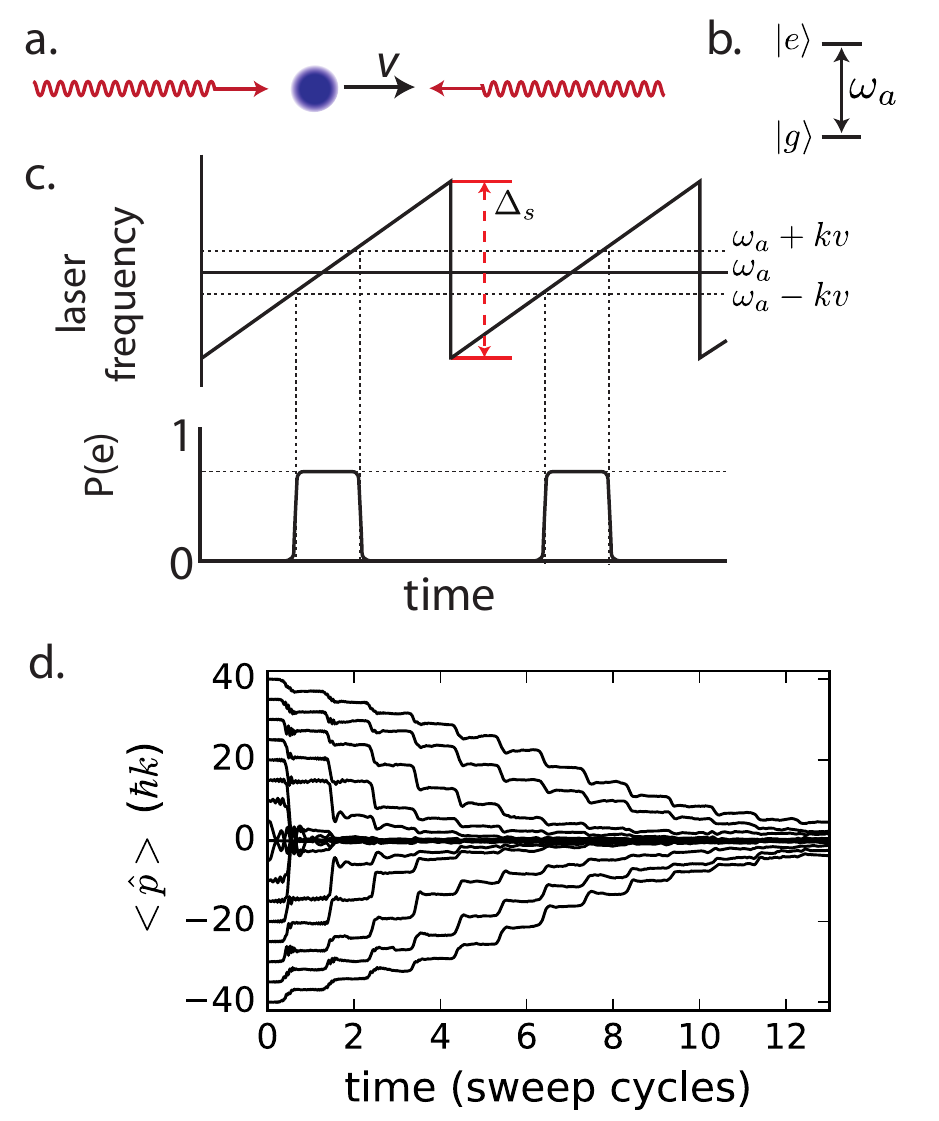}
\caption{Time domain description of cooling mechanism.  (a) A moving atom is illuminated by two counter-propagating laser beams, whose frequencies are modulated with a saw-tooth wave-form. (b) The atom is treated as a two-level system with transition frequency $\omega_a$. (c) Representation of frequencies in lab frame.  Because of Doppler shifts resulting from atomic motion, the co-propagating (counter-propagating) laser beams are resonant with \ket{g} to \ket{e} transition when the laser frequency  $\omega = \omega_a - k v$ ($\omega_a + k v$).  Because $\omega$ increases with time, the counter-propagating laser sweeps over the transition before the co-propagating laser.  If the atom starts in \ket{g}, the counter-propagating laser transfers the atom from \ket{g} to \ket{e} as it sweeps over resonance, and the co-propagating laser transfers the atom back to \ket{g}.  This results in the transfer of two photon recoils of momenta to the atom in the correct direction to slow the atom.     
(d) Simulation of single atom trajectories in momentum space. When the atoms have large momenta, each sweep lowers the momentum by $2 \hbar k$. At lower momenta, multi-photon processes can transfer larger amounts of momentum. }
\label{fig:basics}
\end{figure}

To understand our cooling mechanism in its simplest form, we consider a two-level atom with a long-lived optically excited state \ket{e} and ground state \ket{g} moving in one dimension with velocity~$v$, as shown in Fig.~\ref{fig:basics}a.  Two counterpropagating laser beams with wave-number~$k$ and frequency~$\omega$ are linearly ramped in frequency from below to above the atomic transition frequency~$\omega_a$, with full sweep range~$\Delta_s$. This ramp is repeated to form a sawtooth pattern in time, see Fig.~\ref{fig:basics}c. Each laser interacts with the two-level atom with a Rabi frequency~$\Omega$, which is tuned to be larger than the spontaneous decay rate~$\gamma$ from $\ket{e}$ to $\ket{g}$.  The  sweep range is adjusted such that $\Delta_s> \Omega, 4 k v$. We control the frequency sweep rate $\alpha= (d\omega/dt)$ so that it fulfills the Landau-Zener condition $\alpha \ll \Omega^2$ for adiabatic transfer of atomic populations between ground and excited states. Lastly, the jump in laser frequency at the end of each ramp is considered to be instantaneous, i.e. perfectly diabatic.

%, whose frequency is $\omega_a$. In the lab frame, the lasers are instantaneously at the same frequency $\omega$, which is varied in time in an asymmetrical sawtooth ramp from below to above the atomic transition frequency, seeFig.~\ref{fig:basics}b, starting below the atomic. 

%but is fast enough such that $k v/\alpha \ll 1/\gamma$.    

In the reference frame of the atom, both cooling beams start below resonance with the atomic transition.  Doppler shifts due to the atomic motion cause the beam counter-propagating to the atom's velocity to appear $k v$ higher in frequency and the co-propagating beam to appear  $k v$ lower in frequency.  As the beams sweep upward in frequency, the counter-propagating beam sweeps over resonance first, and adiabatically transfers the atom from \ket{g} to \ket{e}.  Because of the long lifetime of the excited state, the atom remains in \ket{e} until the co-propagating beam sweeps over resonance and adiabatically transfers it back to the ground state.  In this process, the atom has absorbed one photon from the beam propagating against its motion, and emitted a photon into the beam propagating along its motion, resulting in a net momentum transfer of $2 \hbar k$ against its motion.  The laser frequency is then diabatically jumped back to its start frequency such that the atom remains in \ket{g}, and then the process is repeated. 

Similar principles have been explored \cite{PhysRevA.75.011402, 0295-5075-33-4-261,CampbellPorto2014, Metcalf2011Swept, soding1997short} to generate large forces.  In these works, the time-ordering of the left and right-going pulses is fixed by parameters external to the atom.  Here, because the time-ordering of the interaction with the two beams is determined by the atomic velocity, the force always opposes the motion of the atom, and thus causes slowing for atoms moving in either direction.   

Fig.~\ref{fig:basics}d shows simulated trajectories for atoms that begin at a range of different initial momenta using a Monte Carlo wave function trajectory method. Atoms are prepared in a pure quantum state specified by their internal ground state and fixed momentum \cite{dum1992monte}. The average momentum over 50 trajectories is calculated at each time step. At high momentum, we observe that each sweep lowers the momentum of the atom by roughly $2 \hbar k$.   At lower momentum, we find that the change in momentum per sweep can greatly exceed $2 \hbar k$,  as multiple photons are transferred between the two cooling beams.  

%A dressed-state picture (presented in supplemental material) gives some intuition into these multi-photon processes.  

%which we attribute to multi-photon transitions sometimes referred to as Dopplerons \cite{kyrola1977velocity}.

At low momentum, the role of spontaneous emission becomes more important.  As the atom is slowed to near-zero velocity, the Doppler shift becomes small compared to the Rabi frequency, $k v \lesssim \Omega$, and the condition for deterministic time ordering of adiabatic transfers from the two beams no longer holds.  When this occurs, the simple picture of sequential adiabatic transfers to and from \ket{e} becomes invalid, and the probability that the atom is found in \ket{e} at the end of the sweep, which ideally should be 0, becomes appreciable.  While the sawtooth frequency sweep leads to slowing for an atom that is in the ground state at the beginning of a sweep, the effect is reversed for an atom that starts in the excited state, which causes acceleration away from zero velocity.  In the absence of spontaneous emission, the atom would then find itself on a heating trajectory after initial cooling.  The presence of spontaneous emission is therefore critical, as it ensures that the atom preferentially starts in \ket{g} at the beginning of each sweep.   This breaks time reversibility, enabling cooling and phase space compression to occur.

%%%%%%%%%%%%%%%%% Fig 2 %%%%%%%%%%%%%

\begin{figure}[!htb]
\includegraphics[width=3.75in]{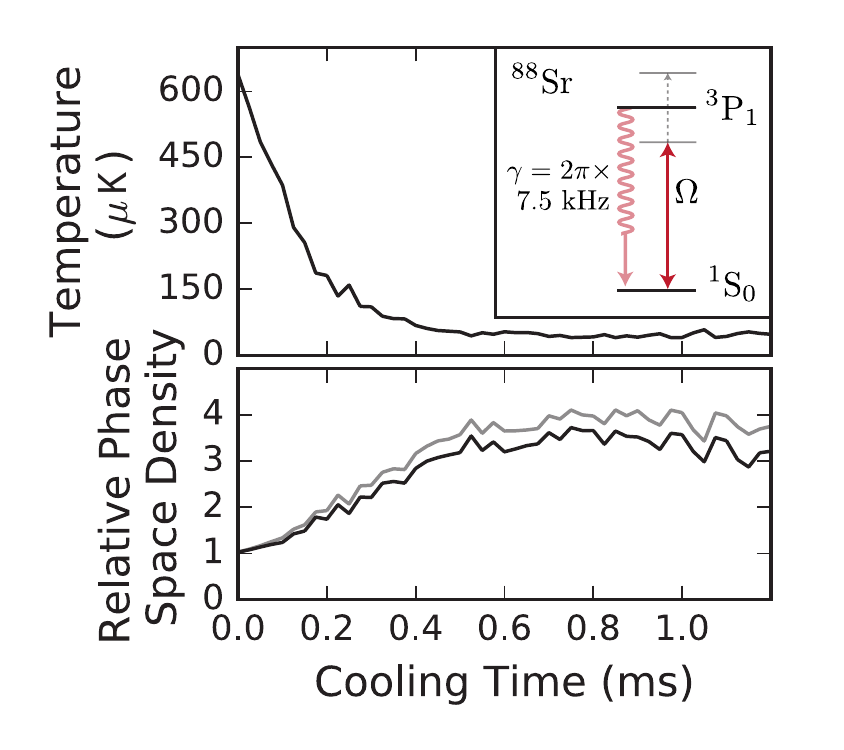}
\caption{Experimental study of an atomic ensemble using one-dimensional cooling by adiabatic transfer.  (a) An ensemble of atoms, precooled to around 600 $\mu$K, is illuminated by frequency-swept counterpropagating beams for a variable amount of time.  The one-dimensional temperature is decreased to a steady-state value of 45 $\mu$K by the cooling lasers.  (b) The phase-space density in one dimension (grey trace) and three dimensions (black trace) is increased during the cooling process. }
\label{fig:cooling}
\end{figure}

We experimentally verify that our mechanism leads to cooling by applying two counter-propagating beams to an ensemble of $^{88}$Sr atoms pre-cooled to roughly 600~$\mu$K (Fig.~\ref{fig:cooling}a). The frequency of the two beams is swept upwards by $\Delta_s= 2 \pi \times 6.6$~MHz every 50~$\mu$s.  The frequency sweep is centered on the resonance of the dipole-forbidden $^1$S$_0$ to $^3$P$_1$ transition, which has a linewidth of $\gamma = 2 \pi \times 7.5$~kHz.  The two beams are linearly polarized in the direction of a 3.3~gauss magnetic field, such that the light-atom interaction can be described as a two-level system with $\ket{g}\equiv \ket{^1\mathrm{S}_0, m_j=0}$ and $\ket{e}\equiv \ket{^3\mathrm{P}_1, m_j=0}$, where $m_j$ labels the magnetic Zeeman sub-level.  

The cooling beams are applied for a time of 50~$\mu$s to 1200~$\mu$s (i.e.\ application of 1 to 24 sweeps) after which the beams are turned off.  The temperature of the atomic cloud is determined from a fluorescence image of the cloud  after free ballistic expansion for 10~ms.  In the direction of beam propagation, we find that for these parameters the atoms are cooled from their initial temperature of roughly 600~$\mu$K to 45~$\mu$K in of order 300~$\mu$s  or 6 sweeps.

%The atoms are allowed to expand freely for 10~ms, after wh fluorescence image to infer the temperature from the spatial distribution of the cloud after the free-flight.  In the direction of beam propagation, we find that for these parameters the atoms are cooled from their initial temperature of roughly 600~$\mu$K to 45~$\mu$K.  

Importantly, we directly observe an increase in phase-space density during the cooling process, and not simply velocity reduction.  We measure atom loss during cooling to be negligible and the relative increase in phase space density is $\rho/\rho_\circ = \Delta x_\circ \Delta v_\circ/(\Delta x \Delta v)$, where $\Delta x$ and $\Delta v$ ($\Delta x_\circ$ and $\Delta v_\circ$) are  the measured cloud size and velocity spread after (before) cooling.  In Fig.~\ref{fig:cooling}b, this quantity is shown as the grey line.  The black line in Fig.~\ref{fig:cooling}b accounts for measured heating in the orthogonal directions (which for simplicity we have assumed to be equal in the two directions).  Note that because the atoms are not confined and the cooling is only applied along one dimension, this increase of phase-space density is much smaller than one would obtain with the same decrease in temperature for three-dimensional cooling in a harmonic trap.

%The fact that phase-space density increases clearly indicates that entropy is removed from the atoms during the cooling process.  At first, this appears problematic for this cooling mechanism since it relies on  absorption and stimulated emission between pairs of lasers, which is a process described completely by unitary dynamics that cannot modify the entropy.  However, as described above, spontaneous emission does start to occur when the unitary dynamics have slowed an atom to a sufficiently low velocity.  The time from the start of cooling to when an atom starts spontaneously emitting photons then serves as a measurement of the initial moment state of the atom.   In this view, an atom’s initial entropy in momentum space is mapped onto entropy contained in the measurement record of when photons are and are not spontaneously emitted.  A conceptually identical concept is described in \cite{PhysRevA.94.043416} in which it was shown that even a single spontaneously emitted photon is sufficient to remove all entropy from an atom during optical pumping between internal levels.  More generally, even the possibility of spontaneous emission without actually emitting a photon in a given realization is sufficient to reduce the entropy of the atomic system.   While a detailed analysis of the entropy transfer is beyond the scope of the work here and will be explored more fully in future work, we believe that the above picture captures the essential physics of how phase space compression is achieved.

The fact that phase-space density increases indicates that entropy is removed during the cooling process.  This may seem problematic for a mechanism that relies heavily on unitary dynamics and stimulated emission. However, as discussed above, the presence of even small amounts of spontaneous emission is critical for breaking time reversibility and enabling phase-space compression.  In related works \cite{PhysRevA.94.043416, price2008single}, it was found that the number of scattered photons required to remove entropy from a system can be quite low (of order one).  

In these protocols, the timing of the emitted photons encodes key information about the initial state of the quantum system.  In an idealized version of our cooling mechanism, the dynamics are completely unitary as the atom is slowed from a high initial momentum.  When it reaches some lower momentum threshold, the simple picture of subsequent adiabatic transfer breaks down and the atom begins to scatter photons.  In this way, the initial velocity of the atom is encoded in the time delay before the atom began to scatter photons, and the atom’s initial entropy in momentum space is mapped onto entropy contained in the measurement record of when photons are and are not spontaneously emitted.  While a detailed analysis of the entropy transfer is beyond the scope of the work here and will be explored more fully in future work, we believe that the above picture captures the essential physics of how phase-space compression is achieved.

%Without spontaneous emission, the dynamics are unitary and reversible, and entropy cannot change. However, at low momentum, the deterministic time ordering of adiabatic transfers from the two beams becomes ambiguous and the probability of spontaneous emission increases, as discussed above. The number of spontaneously scattered photons required to remove the entropy can be quite low (of order one), as Ref.~\cite{PhysRevA.94.043416} found in a related study. If a particular atom does not emit a photon for a long time after the start of cooling, then it must have started at a higher initial momentum. We may think of the unitary dynamics of photon transfer between the counter-propagating beams as removing energy from the system, while the entropy of the initial atomic state is removed by the spontaneously emitted photons.   

%However, as discussed above, the presence of even a single spontaneously emitted photon is sufficient for entropy removal.  The initial atomic momentum state is encoded in the timing of spontaneously emitted photons:  if a particular atom does not emit a photon for a long time after the start of cooling, then it must have started at a higher initial momentum.   We may think of the unitary dynamics of photon transfer between the counter-propagating beams as removing energy from the system, while the entropy of the initial atomic state is encoded in the timing of spontaneously emitted photons.  

 \begin{figure}[!htb]
\includegraphics[width=3.75in]{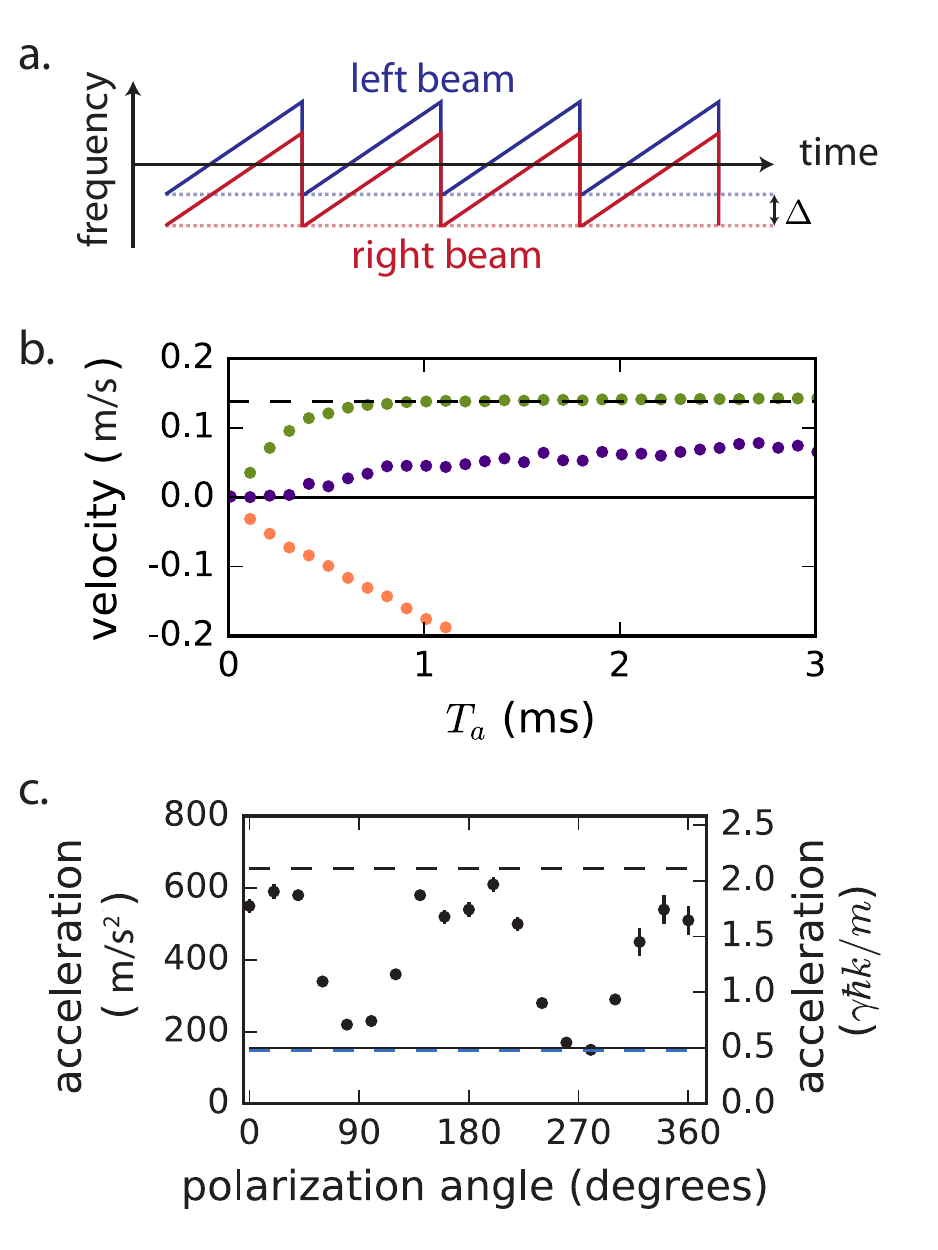}
\caption{Cooling into a moving reference frame.  (a) We apply a frequency offset in the lab reference frame between the two beams, which creates a moving reference frame in which the beams have equal frequency.  (b) If the laser frequency is swept upwards, (green points) the atoms settle into the moving reference frame, which has a velocity indicated by the dashed black line.  If the laser frequency is swept downwards (orange points), the atoms accelerate in the other direction, without cooling.  If the frequency sweep is symmetric (purple points), atoms experience a small acceleration due to radiation-pressure imbalance. (c)  Acceleration versus polarization angle between two beams.  Acceleration is enhanced when the polarizations of the two beams are aligned.  The solid line represents the maximum acceleration for a saturated two-level atom without stimulated emission.  The dashed blue line represents the measured acceleration with a single beam.  Black points represent acceleration with both beams applied.  The dashed black line represents the acceleration that two photon recoils per sweep of the laser frequency would produce. }
\label{fig:versus_time}
\end{figure}

%%%%%%%%%%%%%%%%% Fig 3 %%%%%%%%%%%%%

We now turn to a more detailed characterization of the forces involved in the cooling process. 
Previously, we have considered the case where the instantaneous frequencies of the two beams are equal in the lab frame.  If, however, we introduce a fixed relative offset frequency between the counter-propagating beams (as shown in Fig.~\ref{fig:versus_time}a), then atoms that are stationary in the lab frame will be accelerated until their velocity matches that of the frame $F$ moving at velocity $v_F = \Delta/2k$, in which the two beams appear to have equal instantaneous frequencies.  

%The atoms are prepared nominally at rest in the lab frame by laser cooling in a standard magneto optical trap (MOT) to around 5~$\mu$K.  We then rapidly turn off the three-dimensional trapping beams and zero the magnetic field gradients. Finally, we switch on a small uniform bias magnetic field along $\hat{z}$.

We apply the frequency-swept beams for a variable amount of time $T_a$, then measure the resulting velocity of the atoms.  This is shown in Fig.~\ref{fig:versus_time}b for several sweep configurations. The green points represent the cooling configuration described above, with the laser frequency adiabatically swept from low to high and diabatically jumped from high to low.  The two beams are offset by a detuning $\Delta/2 \pi = 400$~kHz, and are swept by $\Delta_s/2 \pi = 8$~MHz every 33~$\mu$s.  
%The polarizations of the two beams are both oriented along the magnetic field direction $\hat{z}$.  
We observe that the atoms undergo initial acceleration until they reach equilibrium with the velocity of the moving frame $v_F= 0.14$~m/s.  

%.141~m/s.  This is very close to the velocity $v_F= 0.14$~m/s that one would expect from a damping force in the moving frame $F$, with the discrepancy likely due to radiation pressure imbalance between the two beams.  

To rule out the interpretation that the atoms are merely being dragged by the moving standing wave formed by the detuned lasers, we reverse the direction of the sawtooth sweep without changing the relative detuning $\Delta$ of the two beams. The standing wave still moves in the same direction as before, but now the atoms accelerate in the other direction.  While the upwards sweep causes saturation of the atomic velocity at $v_F$, no such saturation is apparent for a downwards sweep.  This confirms that the downwards sweep configuration does not lead to cooling.  

Finally, when we apply a symmetric triangle-ramped frequency sweep to the lasers, the atoms undergo a much smaller acceleration, likely due to incidental residual radiation pressure from an intensity imbalance between the two beams.

%For the upwards sweep, the momentum change imparted by the lasers accelerates the atoms in a direction that reduces the apparent detuning of the two beams in the frame of the atoms.  As this detuning goes to zero, the acceleration mechanism turns off.  For the downwards sweep, the opposite situation occurs, and the acceleration does not turn off until the Doppler shift becomes comparable to the sweep range.  

%By varying the relative polarization of the two beams, we explore how the presence of both beams leads to much larger acceleration than can be achieved with either one ~\ref{fig:versus_time}c).  
%We apply the beams for a time much shorter than it takes the atoms to reach equilibrium velocity in order to infer an initial acceleration.  

Because our mechanism relies on stimulated emission, much larger accelerations can be achieved than would be possible with Doppler cooling on such a narrow transition.  To quantify this acceleration, we apply the beams for a time much shorter than it takes the atoms to reach equilibrium velocity, and measure the resulting change in velocity.  If we apply only one of the two cooling beams, the atoms experience an acceleration consistent with $(\gamma/2) (\hbar k/m)= 155$~m/s$^2$, the expected value for a maximally saturated atom.

%A bias field in the $Z$ direction splits out the $m_j = \pm 1$ Zeeman sublevels by $\pm 7$~MHz.  The leftward travelling beam is always polarized in the $Z$ direction.  The laser frequencies are swept by $\pm 2$~MHz at a repetition rate of 50~kHz, with a detuning $\Delta = 300$~kHz and a Rabi frequency $\Omega = 2\pi \times$~1.3~MHz. By setting the acceleration time $T_{accel}$ to be much less than the timescale on which the atoms reach the velocity of the moving frame, $T_{accel} = 0.1$~ms, we can infer the initial acceleration.   

%
% single-beam accelleration numbers from 6/16.  150 m/s^2 and -160 m/s^2.  Taken with High B field and 0.2 ms acclel time. 

% We observe accelerations of  0.47$m/s^2$ to the left, and 0.50$m/s^2$ to the right.  For a single beam, we find that the shape and direction of the frequency sweep are unimportant.  

We see a far more dramatic effect when we apply both beams at the same time (Fig.~\ref{fig:versus_time}c).  When the polarizations of the two beams are aligned, we observe a maximum acceleration of around 600~m/s$^2$ for a sweep period of $20~\mu$s, a factor of almost 4 above both the observed acceleration from the leftward beam alone and the maximum expected acceleration for a two-level atom without stimulated emission.  This measured acceleration is within ten percent of the value we would expect if each sweep of the laser frequency led to two photon recoils of momentum transfer, though this agreement may partly be due to a cancellation between the effects of imperfect adiabatic transfers and multi-photon processes.   
When the polarizations are made orthogonal, the acceleration returns to near the single-beam value, as only one of the beams interacts with $\ket{e}$.

%, as one beam interacts with $\ket{e}$, while the other beam interacts with the other two zeeman states $\ket{^3\mathrm{P}_1, m_j=\pm1}$ such that atoms are no longer returned to the ground state by stimulated emission.  

%In contrast to Doppler cooling, the presence of the laser co-propagating with atomic motion enhances slowing.  Interestingly, the radiation forces associated with this laser serve to pull the atom toward toward the laser source.  

%%%%%%%%%%%%%%%%% Fig 4 %%%%%%%%%%%%%

\begin{figure}[!htb]
\includegraphics[width=3.375in]{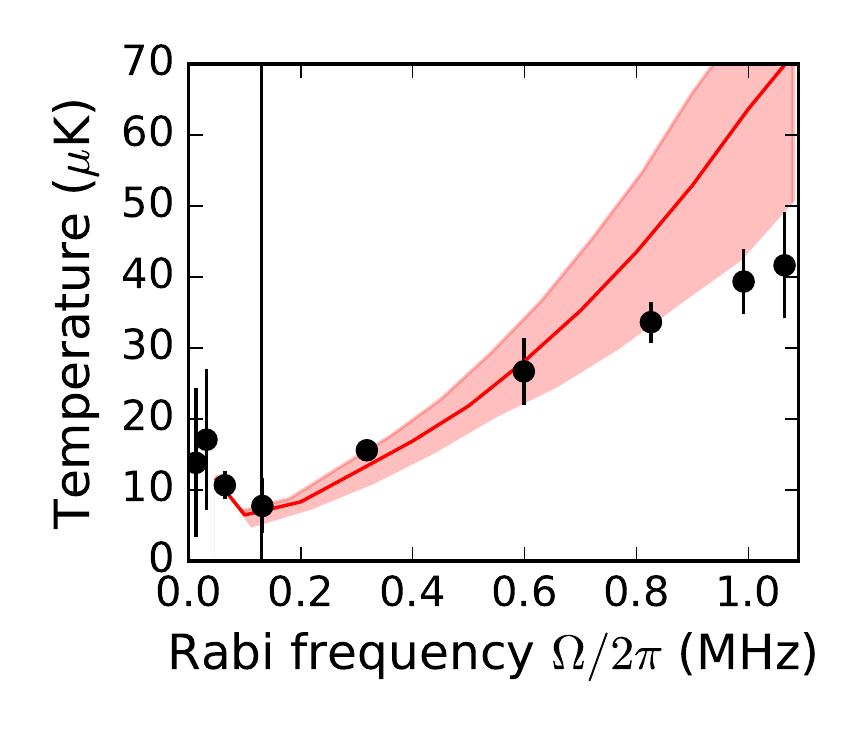}
\caption{Temperature versus Rabi frequency for a sweep range of 7~MHz and repetition rate of 15~kHz.  Red line represents the results of our simulations, and black points are experimental data.  The red band indicates uncertainty of prediction due to statistical uncertainty in simulation results, experimental calibration of $\Omega$, and variations in Rabi frequency during the sweep.  The vertical line at 130~kHz represents the approximate point at which we expect adiabaticity to break down: $\Omega^2 = \alpha$.  
%The blue band represents a temperature of $\hbar \Omega/kB$, with width representing our uncertainty in $\omega$.  This prediction for the temperature is quite close to our measured values, though our theory predicts that this is not the correct scaling.  
}
\label{fig:rabi}
\end{figure}

An analytic or intuitive prediction of the final temperature has proved difficult, especially in the presence of multi-photon processes.  We compare our experimentally measured temperature to simulation in Fig.~\ref{fig:rabi}.  

The simulation used is a Monte Carlo wave function approach. While the external electric field associated with the cooling lasers is treated classically, the internal and external atomic degrees of freedom are treated fully quantum mechanically. The wave function contains ground and excited internal states in addition to a discrete family of external momentum states, which are equally spaced in units of $\hbar k$ up to a maximum cutoff. The unitary dynamics are generated by the single-particle Hamiltonian

\begin{align}
    \hat{H} & = 
        \frac{\hat{p}^2}{2m}
        + \frac{1}{2} \hbar \omega_a \hat{\sigma}^z \notag \\
        & \quad + \hbar \Omega \cos ( k \hat{z} )
            \left(
            \hat{\sigma}^+ e^{-i \eta(t)} + \text{h.c.}
            \right),
\end{align}

\noindent where the instantaneous phase of the applied coherent field $\eta(t)$ is the time integrated instantaneous frequency, with the frequency ramped in the sawtooth pattern previously described. Because the full atom and field system is an open quantum system due to the presence of spontaneous emission, a decay operator proportional to the spontaneous emission rate is also included to fully simulate the associated quantum master equation.

Both experiment and simulation show a minimum temperature as a function of Rabi frequency.  At low Rabi frequency, the Landau-Zener condition $\alpha \ll \Omega^2$ breaks down, leading to inefficient adiabatic transfers.  For the parameters used in (Fig.~\ref{fig:rabi}), $\alpha = \Omega^2$ when $\Omega = 2 \pi \times 130$~kHz, roughly the point at which temperature is minimized in both the simulation and experimental results.  

At larger $\Omega$, both the experimentally measured and simulated temperatures rise with $\Omega$, though the predicted and observed values appear to disagree at large $\Omega$.  
Several experimental factors could lead to this discrepancy.  Because an acousto-optic modulator (AOM) is used to sweep the frequency of the cooling lasers, there are significant power variations during the sweep.  Including this effect in our simulation resulted in lower temperatures at high Rabi frequency, and this effect can explain much of the disagreement (as indicated by the red band in Fig.~\ref{fig:rabi}).  Furthermore, imperfect laser polarization and finite bias field may lead to cooling on the transitions to $^3$P$_1$, $m_j = \pm 1$, especially at high Rabi frequency.

In conclusion, we have demonstrated a novel cooling mechanism in which large accelerations may be achieved even on optical transitions with low scattering rates.  This mechanism may find application to the cooling of molecules, and can improve the performance of atomic cooling using dipole-forbidden optical transitions.  From a technical perspective, this cooling method has the advantage of low sensitivity to long-term laser frequency drifts and to perturbations of the atomic transition frequency, for example to those encountered while loading into a deep optical dipole trap \cite{doi:10.1143/JPSJ.68.2479}.

%The narrow linewidth and low scattering rate from these dipole-forbidden transitions leads to challenges for standard Doppler cooling.  The low scattering rate from even a fully saturated atom limits the maximum achievable force exerted on the atom, resulting in a low slowing rate and a low capture velocity in a trap.  Further, in order to take advantage of the narrow linewidth of the transition, a laser with long-term stability comparable to the (power broadened) transition linewidth is required. Similarly, effective cooling is prevented by relatively small frequency shifts to the cooling transition, which complicates the cooling of atoms in deep optical dipole traps \cite{doi:10.1143/JPSJ.68.2479}.

%Further, in species with hyperfine structure and narrow linewidth transitions (such as $^{87}$Sr), atoms can become stuck in unconfined ground states.  This requires additional ``stirring" beams, increasing experimental complexity \cite{PhysRevLett.90.113002}.  
%The cooling mechanism that we present here mitigates all of these issues.  

Extensions and modifications to this cooling mechanism may include the use of circularly polarized beams and magnetic field gradients to form a magneto-optical trap (also observed but to be described elsewhere), near ground state cooling of tightly confined atoms, and application to species without narrow linewidth optical transitions by using Raman transitions.  Furthermore, this method could be employed to accelerate, deccelerate, or otherwise control atomic beams or ensembles with strong stimulated forces.

\section{ACKNOWLEDGMENTS}\label{ack}
  
All authors acknowledge financial support from DARPA QuASAR, ARO, NSF PFC, and NIST. J.R.K.C. acknowledges financial support from NSF GRFP. This work is supported by the National Science Foundation under Grant Number 1125844.

\bibliographystyle{apsrev4-1}
\bibliography{ThompsonLab.bib}

\begin{thebibliography}{24}%
\makeatletter
\providecommand \@ifxundefined [1]{%
 \@ifx{#1\undefined}
}%
\providecommand \@ifnum [1]{%
 \ifnum #1\expandafter \@firstoftwo
 \else \expandafter \@secondoftwo
 \fi
}%
\providecommand \@ifx [1]{%
 \ifx #1\expandafter \@firstoftwo
 \else \expandafter \@secondoftwo
 \fi
}%
\providecommand \natexlab [1]{#1}%
\providecommand \enquote  [1]{``#1''}%
\providecommand \bibnamefont  [1]{#1}%
\providecommand \bibfnamefont [1]{#1}%
\providecommand \citenamefont [1]{#1}%
\providecommand \href@noop [0]{\@secondoftwo}%
\providecommand \href [0]{\begingroup \@sanitize@url \@href}%
\providecommand \@href[1]{\@@startlink{#1}\@@href}%
\providecommand \@@href[1]{\endgroup#1\@@endlink}%
\providecommand \@sanitize@url [0]{\catcode `\\12\catcode `\$12\catcode
  `\&12\catcode `\#12\catcode `\^12\catcode `\_12\catcode `\%12\relax}%
\providecommand \@@startlink[1]{}%
\providecommand \@@endlink[0]{}%
\providecommand \url  [0]{\begingroup\@sanitize@url \@url }%
\providecommand \@url [1]{\endgroup\@href {#1}{\urlprefix }}%
\providecommand \urlprefix  [0]{URL }%
\providecommand \Eprint [0]{\href }%
\providecommand \doibase [0]{http://dx.doi.org/}%
\providecommand \selectlanguage [0]{\@gobble}%
\providecommand \bibinfo  [0]{\@secondoftwo}%
\providecommand \bibfield  [0]{\@secondoftwo}%
\providecommand \translation [1]{[#1]}%
\providecommand \BibitemOpen [0]{}%
\providecommand \bibitemStop [0]{}%
\providecommand \bibitemNoStop [0]{.\EOS\space}%
\providecommand \EOS [0]{\spacefactor3000\relax}%
\providecommand \BibitemShut  [1]{\csname bibitem#1\endcsname}%
\let\auto@bib@innerbib\@empty
%</preamble>
\bibitem [{\citenamefont {Chu}\ \emph {et~al.}(1986)\citenamefont {Chu},
  \citenamefont {Bjorkholm}, \citenamefont {Ashkin},\ and\ \citenamefont
  {Cable}}]{chu1986experimental}%
  \BibitemOpen
  \bibfield  {author} {\bibinfo {author} {\bibfnamefont {S.}~\bibnamefont
  {Chu}}, \bibinfo {author} {\bibfnamefont {J.}~\bibnamefont {Bjorkholm}},
  \bibinfo {author} {\bibfnamefont {A.}~\bibnamefont {Ashkin}}, \ and\ \bibinfo
  {author} {\bibfnamefont {A.}~\bibnamefont {Cable}},\ }\href@noop {}
  {\bibfield  {journal} {\bibinfo  {journal} {Physical Review Letters}\
  }\textbf {\bibinfo {volume} {57}},\ \bibinfo {pages} {314} (\bibinfo {year}
  {1986})}\BibitemShut {NoStop}%
\bibitem [{\citenamefont {Diedrich}\ \emph {et~al.}(1989)\citenamefont
  {Diedrich}, \citenamefont {Bergquist}, \citenamefont {Itano},\ and\
  \citenamefont {Wineland}}]{diedrich1989laser}%
  \BibitemOpen
  \bibfield  {author} {\bibinfo {author} {\bibfnamefont {F.}~\bibnamefont
  {Diedrich}}, \bibinfo {author} {\bibfnamefont {J.}~\bibnamefont {Bergquist}},
  \bibinfo {author} {\bibfnamefont {W.~M.}\ \bibnamefont {Itano}}, \ and\
  \bibinfo {author} {\bibfnamefont {D.}~\bibnamefont {Wineland}},\ }\href@noop
  {} {\bibfield  {journal} {\bibinfo  {journal} {Physical Review Letters}\
  }\textbf {\bibinfo {volume} {62}},\ \bibinfo {pages} {403} (\bibinfo {year}
  {1989})}\BibitemShut {NoStop}%
\bibitem [{\citenamefont {Chan}\ \emph {et~al.}(2011)\citenamefont {Chan},
  \citenamefont {Alegre}, \citenamefont {Safavi-Naeini}, \citenamefont {Hill},
  \citenamefont {Krause}, \citenamefont {Gr{\"o}blacher}, \citenamefont
  {Aspelmeyer},\ and\ \citenamefont {Painter}}]{chan2011laser}%
  \BibitemOpen
  \bibfield  {author} {\bibinfo {author} {\bibfnamefont {J.}~\bibnamefont
  {Chan}}, \bibinfo {author} {\bibfnamefont {T.~M.}\ \bibnamefont {Alegre}},
  \bibinfo {author} {\bibfnamefont {A.~H.}\ \bibnamefont {Safavi-Naeini}},
  \bibinfo {author} {\bibfnamefont {J.~T.}\ \bibnamefont {Hill}}, \bibinfo
  {author} {\bibfnamefont {A.}~\bibnamefont {Krause}}, \bibinfo {author}
  {\bibfnamefont {S.}~\bibnamefont {Gr{\"o}blacher}}, \bibinfo {author}
  {\bibfnamefont {M.}~\bibnamefont {Aspelmeyer}}, \ and\ \bibinfo {author}
  {\bibfnamefont {O.}~\bibnamefont {Painter}},\ }\href@noop {} {\bibfield
  {journal} {\bibinfo  {journal} {Nature}\ }\textbf {\bibinfo {volume} {478}},\
  \bibinfo {pages} {89} (\bibinfo {year} {2011})}\BibitemShut {NoStop}%
\bibitem [{\citenamefont {Katori}\ \emph
  {et~al.}(1999{\natexlab{a}})\citenamefont {Katori}, \citenamefont {Ido},
  \citenamefont {Isoya},\ and\ \citenamefont
  {Kuwata-Gonokami}}]{PhysRevLett.82.1116}%
  \BibitemOpen
  \bibfield  {author} {\bibinfo {author} {\bibfnamefont {H.}~\bibnamefont
  {Katori}}, \bibinfo {author} {\bibfnamefont {T.}~\bibnamefont {Ido}},
  \bibinfo {author} {\bibfnamefont {Y.}~\bibnamefont {Isoya}}, \ and\ \bibinfo
  {author} {\bibfnamefont {M.}~\bibnamefont {Kuwata-Gonokami}},\ }\href
  {\doibase 10.1103/PhysRevLett.82.1116} {\bibfield  {journal} {\bibinfo
  {journal} {Phys. Rev. Lett.}\ }\textbf {\bibinfo {volume} {82}},\ \bibinfo
  {pages} {1116} (\bibinfo {year} {1999}{\natexlab{a}})}\BibitemShut {NoStop}%
\bibitem [{\citenamefont {Vogel}\ \emph {et~al.}(1999)\citenamefont {Vogel},
  \citenamefont {Dinneen}, \citenamefont {Gallagher},\ and\ \citenamefont
  {Hall}}]{vogel1999narrow}%
  \BibitemOpen
  \bibfield  {author} {\bibinfo {author} {\bibfnamefont {K.~R.}\ \bibnamefont
  {Vogel}}, \bibinfo {author} {\bibfnamefont {T.~P.}\ \bibnamefont {Dinneen}},
  \bibinfo {author} {\bibfnamefont {A.}~\bibnamefont {Gallagher}}, \ and\
  \bibinfo {author} {\bibfnamefont {J.~L.}\ \bibnamefont {Hall}},\ }\href@noop
  {} {\bibfield  {journal} {\bibinfo  {journal} {IEEE Transactions on
  Instrumentation and Measurement}\ }\textbf {\bibinfo {volume} {48}},\
  \bibinfo {pages} {618} (\bibinfo {year} {1999})}\BibitemShut {NoStop}%
\bibitem [{\citenamefont {Stellmer}\ \emph {et~al.}(2013)\citenamefont
  {Stellmer}, \citenamefont {Pasquiou}, \citenamefont {Grimm},\ and\
  \citenamefont {Schreck}}]{PhysRevLett.110.263003}%
  \BibitemOpen
  \bibfield  {author} {\bibinfo {author} {\bibfnamefont {S.}~\bibnamefont
  {Stellmer}}, \bibinfo {author} {\bibfnamefont {B.}~\bibnamefont {Pasquiou}},
  \bibinfo {author} {\bibfnamefont {R.}~\bibnamefont {Grimm}}, \ and\ \bibinfo
  {author} {\bibfnamefont {F.}~\bibnamefont {Schreck}},\ }\href {\doibase
  10.1103/PhysRevLett.110.263003} {\bibfield  {journal} {\bibinfo  {journal}
  {Phys. Rev. Lett.}\ }\textbf {\bibinfo {volume} {110}},\ \bibinfo {pages}
  {263003} (\bibinfo {year} {2013})}\BibitemShut {NoStop}%
\bibitem [{\citenamefont {Duarte}\ \emph {et~al.}(2011)\citenamefont {Duarte},
  \citenamefont {Hart}, \citenamefont {Hitchcock}, \citenamefont {Corcovilos},
  \citenamefont {Yang}, \citenamefont {Reed},\ and\ \citenamefont
  {Hulet}}]{PhysRevA.84.061406}%
  \BibitemOpen
  \bibfield  {author} {\bibinfo {author} {\bibfnamefont {P.~M.}\ \bibnamefont
  {Duarte}}, \bibinfo {author} {\bibfnamefont {R.~A.}\ \bibnamefont {Hart}},
  \bibinfo {author} {\bibfnamefont {J.~M.}\ \bibnamefont {Hitchcock}}, \bibinfo
  {author} {\bibfnamefont {T.~A.}\ \bibnamefont {Corcovilos}}, \bibinfo
  {author} {\bibfnamefont {T.-L.}\ \bibnamefont {Yang}}, \bibinfo {author}
  {\bibfnamefont {A.}~\bibnamefont {Reed}}, \ and\ \bibinfo {author}
  {\bibfnamefont {R.~G.}\ \bibnamefont {Hulet}},\ }\href {\doibase
  10.1103/PhysRevA.84.061406} {\bibfield  {journal} {\bibinfo  {journal} {Phys.
  Rev. A}\ }\textbf {\bibinfo {volume} {84}},\ \bibinfo {pages} {061406}
  (\bibinfo {year} {2011})}\BibitemShut {NoStop}%
\bibitem [{\citenamefont {Ido}\ \emph {et~al.}(2000)\citenamefont {Ido},
  \citenamefont {Isoya},\ and\ \citenamefont {Katori}}]{PhysRevA.61.061403}%
  \BibitemOpen
  \bibfield  {author} {\bibinfo {author} {\bibfnamefont {T.}~\bibnamefont
  {Ido}}, \bibinfo {author} {\bibfnamefont {Y.}~\bibnamefont {Isoya}}, \ and\
  \bibinfo {author} {\bibfnamefont {H.}~\bibnamefont {Katori}},\ }\href
  {\doibase 10.1103/PhysRevA.61.061403} {\bibfield  {journal} {\bibinfo
  {journal} {Phys. Rev. A}\ }\textbf {\bibinfo {volume} {61}},\ \bibinfo
  {pages} {061403} (\bibinfo {year} {2000})}\BibitemShut {NoStop}%
\bibitem [{\citenamefont {Loftus}\ \emph {et~al.}(2004)\citenamefont {Loftus},
  \citenamefont {Ido}, \citenamefont {Ludlow}, \citenamefont {Boyd},\ and\
  \citenamefont {Ye}}]{PhysRevLett.93.073003}%
  \BibitemOpen
  \bibfield  {author} {\bibinfo {author} {\bibfnamefont {T.~H.}\ \bibnamefont
  {Loftus}}, \bibinfo {author} {\bibfnamefont {T.}~\bibnamefont {Ido}},
  \bibinfo {author} {\bibfnamefont {A.~D.}\ \bibnamefont {Ludlow}}, \bibinfo
  {author} {\bibfnamefont {M.~M.}\ \bibnamefont {Boyd}}, \ and\ \bibinfo
  {author} {\bibfnamefont {J.}~\bibnamefont {Ye}},\ }\href {\doibase
  10.1103/PhysRevLett.93.073003} {\bibfield  {journal} {\bibinfo  {journal}
  {Phys. Rev. Lett.}\ }\textbf {\bibinfo {volume} {93}},\ \bibinfo {pages}
  {073003} (\bibinfo {year} {2004})}\BibitemShut {NoStop}%
\bibitem [{\citenamefont {Shuman}\ \emph {et~al.}(2010)\citenamefont {Shuman},
  \citenamefont {Barry},\ and\ \citenamefont {DeMille}}]{shuman2010laser}%
  \BibitemOpen
  \bibfield  {author} {\bibinfo {author} {\bibfnamefont {E.~S.}\ \bibnamefont
  {Shuman}}, \bibinfo {author} {\bibfnamefont {J.~F.}\ \bibnamefont {Barry}}, \
  and\ \bibinfo {author} {\bibfnamefont {D.}~\bibnamefont {DeMille}},\
  }\href@noop {} {\bibfield  {journal} {\bibinfo  {journal} {Nature}\ }\textbf
  {\bibinfo {volume} {467}},\ \bibinfo {pages} {820} (\bibinfo {year}
  {2010})}\BibitemShut {NoStop}%
\bibitem [{\citenamefont {Hummon}\ \emph {et~al.}(2013)\citenamefont {Hummon},
  \citenamefont {Yeo}, \citenamefont {Stuhl}, \citenamefont {Collopy},
  \citenamefont {Xia},\ and\ \citenamefont {Ye}}]{hummon20132d}%
  \BibitemOpen
  \bibfield  {author} {\bibinfo {author} {\bibfnamefont {M.~T.}\ \bibnamefont
  {Hummon}}, \bibinfo {author} {\bibfnamefont {M.}~\bibnamefont {Yeo}},
  \bibinfo {author} {\bibfnamefont {B.~K.}\ \bibnamefont {Stuhl}}, \bibinfo
  {author} {\bibfnamefont {A.~L.}\ \bibnamefont {Collopy}}, \bibinfo {author}
  {\bibfnamefont {Y.}~\bibnamefont {Xia}}, \ and\ \bibinfo {author}
  {\bibfnamefont {J.}~\bibnamefont {Ye}},\ }\href@noop {} {\bibfield  {journal}
  {\bibinfo  {journal} {Physical Review Letters}\ }\textbf {\bibinfo {volume}
  {110}},\ \bibinfo {pages} {143001} (\bibinfo {year} {2013})}\BibitemShut
  {NoStop}%
\bibitem [{\citenamefont {Katori}\ \emph
  {et~al.}(1999{\natexlab{b}})\citenamefont {Katori}, \citenamefont {Ido},\
  and\ \citenamefont {Kuwata-Gonokami}}]{doi:10.1143/JPSJ.68.2479}%
  \BibitemOpen
  \bibfield  {author} {\bibinfo {author} {\bibfnamefont {H.}~\bibnamefont
  {Katori}}, \bibinfo {author} {\bibfnamefont {T.}~\bibnamefont {Ido}}, \ and\
  \bibinfo {author} {\bibfnamefont {M.}~\bibnamefont {Kuwata-Gonokami}},\
  }\href {\doibase 10.1143/JPSJ.68.2479} {\bibfield  {journal} {\bibinfo
  {journal} {Journal of the Physical Society of Japan}\ }\textbf {\bibinfo
  {volume} {68}},\ \bibinfo {pages} {2479} (\bibinfo {year}
  {1999}{\natexlab{b}})}\BibitemShut {NoStop}%
\bibitem [{\citenamefont {Collopy}\ \emph {et~al.}(2015)\citenamefont
  {Collopy}, \citenamefont {Hummon}, \citenamefont {Yeo}, \citenamefont {Yan},\
  and\ \citenamefont {Ye}}]{collopy2015prospects}%
  \BibitemOpen
  \bibfield  {author} {\bibinfo {author} {\bibfnamefont {A.~L.}\ \bibnamefont
  {Collopy}}, \bibinfo {author} {\bibfnamefont {M.~T.}\ \bibnamefont {Hummon}},
  \bibinfo {author} {\bibfnamefont {M.}~\bibnamefont {Yeo}}, \bibinfo {author}
  {\bibfnamefont {B.}~\bibnamefont {Yan}}, \ and\ \bibinfo {author}
  {\bibfnamefont {J.}~\bibnamefont {Ye}},\ }\href@noop {} {\bibfield  {journal}
  {\bibinfo  {journal} {New Journal of Physics}\ }\textbf {\bibinfo {volume}
  {17}},\ \bibinfo {pages} {055008} (\bibinfo {year} {2015})}\BibitemShut
  {NoStop}%
\bibitem [{\citenamefont {Kobayashi}\ \emph {et~al.}(2014)\citenamefont
  {Kobayashi}, \citenamefont {Aikawa}, \citenamefont {Oasa},\ and\
  \citenamefont {Inouye}}]{kobayashi2014prospects}%
  \BibitemOpen
  \bibfield  {author} {\bibinfo {author} {\bibfnamefont {J.}~\bibnamefont
  {Kobayashi}}, \bibinfo {author} {\bibfnamefont {K.}~\bibnamefont {Aikawa}},
  \bibinfo {author} {\bibfnamefont {K.}~\bibnamefont {Oasa}}, \ and\ \bibinfo
  {author} {\bibfnamefont {S.}~\bibnamefont {Inouye}},\ }\href@noop {}
  {\bibfield  {journal} {\bibinfo  {journal} {Physical Review A}\ }\textbf
  {\bibinfo {volume} {89}},\ \bibinfo {pages} {021401} (\bibinfo {year}
  {2014})}\BibitemShut {NoStop}%
\bibitem [{\citenamefont {Metcalf}(2008)}]{metcalf2008entropy}%
  \BibitemOpen
  \bibfield  {author} {\bibinfo {author} {\bibfnamefont {H.}~\bibnamefont
  {Metcalf}},\ }\href@noop {} {\bibfield  {journal} {\bibinfo  {journal}
  {Physical Review A}\ }\textbf {\bibinfo {volume} {77}},\ \bibinfo {pages}
  {061401} (\bibinfo {year} {2008})}\BibitemShut {NoStop}%
\bibitem [{\citenamefont {Corder}\ \emph {et~al.}(2015)\citenamefont {Corder},
  \citenamefont {Arnold},\ and\ \citenamefont {Metcalf}}]{corder2015laser}%
  \BibitemOpen
  \bibfield  {author} {\bibinfo {author} {\bibfnamefont {C.}~\bibnamefont
  {Corder}}, \bibinfo {author} {\bibfnamefont {B.}~\bibnamefont {Arnold}}, \
  and\ \bibinfo {author} {\bibfnamefont {H.}~\bibnamefont {Metcalf}},\
  }\href@noop {} {\bibfield  {journal} {\bibinfo  {journal} {Physical Review
  Letters}\ }\textbf {\bibinfo {volume} {114}},\ \bibinfo {pages} {043002}
  (\bibinfo {year} {2015})}\BibitemShut {NoStop}%
\bibitem [{\citenamefont {Rochester}\ \emph {et~al.}(2016)\citenamefont
  {Rochester}, \citenamefont {Szyma\ifmmode~\acute{n}\else \'{n}\fi{}ski},
  \citenamefont {Raizen}, \citenamefont {Pustelny}, \citenamefont {Auzinsh},\
  and\ \citenamefont {Budker}}]{PhysRevA.94.043416}%
  \BibitemOpen
  \bibfield  {author} {\bibinfo {author} {\bibfnamefont {S.~M.}\ \bibnamefont
  {Rochester}}, \bibinfo {author} {\bibfnamefont {K.}~\bibnamefont
  {Szyma\ifmmode~\acute{n}\else \'{n}\fi{}ski}}, \bibinfo {author}
  {\bibfnamefont {M.}~\bibnamefont {Raizen}}, \bibinfo {author} {\bibfnamefont
  {S.}~\bibnamefont {Pustelny}}, \bibinfo {author} {\bibfnamefont
  {M.}~\bibnamefont {Auzinsh}}, \ and\ \bibinfo {author} {\bibfnamefont
  {D.}~\bibnamefont {Budker}},\ }\href {\doibase 10.1103/PhysRevA.94.043416}
  {\bibfield  {journal} {\bibinfo  {journal} {Phys. Rev. A}\ }\textbf {\bibinfo
  {volume} {94}},\ \bibinfo {pages} {043416} (\bibinfo {year}
  {2016})}\BibitemShut {NoStop}%
\bibitem [{\citenamefont {Price}\ \emph {et~al.}(2008)\citenamefont {Price},
  \citenamefont {Bannerman}, \citenamefont {Viering}, \citenamefont
  {Narevicius},\ and\ \citenamefont {Raizen}}]{price2008single}%
  \BibitemOpen
  \bibfield  {author} {\bibinfo {author} {\bibfnamefont {G.~N.}\ \bibnamefont
  {Price}}, \bibinfo {author} {\bibfnamefont {S.~T.}\ \bibnamefont
  {Bannerman}}, \bibinfo {author} {\bibfnamefont {K.}~\bibnamefont {Viering}},
  \bibinfo {author} {\bibfnamefont {E.}~\bibnamefont {Narevicius}}, \ and\
  \bibinfo {author} {\bibfnamefont {M.~G.}\ \bibnamefont {Raizen}},\
  }\href@noop {} {\bibfield  {journal} {\bibinfo  {journal} {Physical Review
  Letters}\ }\textbf {\bibinfo {volume} {100}},\ \bibinfo {pages} {093004}
  (\bibinfo {year} {2008})}\BibitemShut {NoStop}%
\bibitem [{\citenamefont {Miao}\ \emph {et~al.}(2007)\citenamefont {Miao},
  \citenamefont {Wertz}, \citenamefont {Cohen},\ and\ \citenamefont
  {Metcalf}}]{PhysRevA.75.011402}%
  \BibitemOpen
  \bibfield  {author} {\bibinfo {author} {\bibfnamefont {X.}~\bibnamefont
  {Miao}}, \bibinfo {author} {\bibfnamefont {E.}~\bibnamefont {Wertz}},
  \bibinfo {author} {\bibfnamefont {M.~G.}\ \bibnamefont {Cohen}}, \ and\
  \bibinfo {author} {\bibfnamefont {H.}~\bibnamefont {Metcalf}},\ }\href
  {\doibase 10.1103/PhysRevA.75.011402} {\bibfield  {journal} {\bibinfo
  {journal} {Phys. Rev. A}\ }\textbf {\bibinfo {volume} {75}},\ \bibinfo
  {pages} {011402} (\bibinfo {year} {2007})}\BibitemShut {NoStop}%
\bibitem [{\citenamefont {Nölle}\ \emph {et~al.}(1996)\citenamefont {Nölle},
  \citenamefont {Nölle}, \citenamefont {Schmand},\ and\ \citenamefont
  {Andrä}}]{0295-5075-33-4-261}%
  \BibitemOpen
  \bibfield  {author} {\bibinfo {author} {\bibfnamefont {B.}~\bibnamefont
  {Nölle}}, \bibinfo {author} {\bibfnamefont {H.}~\bibnamefont {Nölle}},
  \bibinfo {author} {\bibfnamefont {J.}~\bibnamefont {Schmand}}, \ and\
  \bibinfo {author} {\bibfnamefont {H.~J.}\ \bibnamefont {Andrä}},\ }\href
  {http://stacks.iop.org/0295-5075/33/i=4/a=261} {\bibfield  {journal}
  {\bibinfo  {journal} {EPL (Europhysics Letters)}\ }\textbf {\bibinfo {volume}
  {33}},\ \bibinfo {pages} {261} (\bibinfo {year} {1996})}\BibitemShut
  {NoStop}%
\bibitem [{\citenamefont {Jayich}\ \emph {et~al.}(2014)\citenamefont {Jayich},
  \citenamefont {Vutha}, \citenamefont {Hummon}, \citenamefont {Porto},\ and\
  \citenamefont {Campbell}}]{CampbellPorto2014}%
  \BibitemOpen
  \bibfield  {author} {\bibinfo {author} {\bibfnamefont {A.~M.}\ \bibnamefont
  {Jayich}}, \bibinfo {author} {\bibfnamefont {A.~C.}\ \bibnamefont {Vutha}},
  \bibinfo {author} {\bibfnamefont {M.~T.}\ \bibnamefont {Hummon}}, \bibinfo
  {author} {\bibfnamefont {J.~V.}\ \bibnamefont {Porto}}, \ and\ \bibinfo
  {author} {\bibfnamefont {W.~C.}\ \bibnamefont {Campbell}},\ }\href {\doibase
  10.1103/PhysRevA.89.023425} {\bibfield  {journal} {\bibinfo  {journal} {Phys.
  Rev. A}\ }\textbf {\bibinfo {volume} {89}},\ \bibinfo {pages} {023425}
  (\bibinfo {year} {2014})}\BibitemShut {NoStop}%
\bibitem [{\citenamefont {Stack}\ \emph {et~al.}(2011)\citenamefont {Stack},
  \citenamefont {Elgin}, \citenamefont {Anisimov},\ and\ \citenamefont
  {Metcalf}}]{Metcalf2011Swept}%
  \BibitemOpen
  \bibfield  {author} {\bibinfo {author} {\bibfnamefont {D.}~\bibnamefont
  {Stack}}, \bibinfo {author} {\bibfnamefont {J.}~\bibnamefont {Elgin}},
  \bibinfo {author} {\bibfnamefont {P.~M.}\ \bibnamefont {Anisimov}}, \ and\
  \bibinfo {author} {\bibfnamefont {H.}~\bibnamefont {Metcalf}},\ }\href
  {\doibase 10.1103/PhysRevA.84.013420} {\bibfield  {journal} {\bibinfo
  {journal} {Phys. Rev. A}\ }\textbf {\bibinfo {volume} {84}},\ \bibinfo
  {pages} {013420} (\bibinfo {year} {2011})}\BibitemShut {NoStop}%
\bibitem [{\citenamefont {S{\"o}ding}\ \emph {et~al.}(1997)\citenamefont
  {S{\"o}ding}, \citenamefont {Grimm}, \citenamefont {Ovchinnikov},
  \citenamefont {Bouyer},\ and\ \citenamefont {Salomon}}]{soding1997short}%
  \BibitemOpen
  \bibfield  {author} {\bibinfo {author} {\bibfnamefont {J.}~\bibnamefont
  {S{\"o}ding}}, \bibinfo {author} {\bibfnamefont {R.}~\bibnamefont {Grimm}},
  \bibinfo {author} {\bibfnamefont {Y.~B.}\ \bibnamefont {Ovchinnikov}},
  \bibinfo {author} {\bibfnamefont {P.}~\bibnamefont {Bouyer}}, \ and\ \bibinfo
  {author} {\bibfnamefont {C.}~\bibnamefont {Salomon}},\ }\href@noop {}
  {\bibfield  {journal} {\bibinfo  {journal} {Physical Review Letters}\
  }\textbf {\bibinfo {volume} {78}},\ \bibinfo {pages} {1420} (\bibinfo {year}
  {1997})}\BibitemShut {NoStop}%
\bibitem [{\citenamefont {Dum}\ \emph {et~al.}(1992)\citenamefont {Dum},
  \citenamefont {Zoller},\ and\ \citenamefont {Ritsch}}]{dum1992monte}%
  \BibitemOpen
  \bibfield  {author} {\bibinfo {author} {\bibfnamefont {R.}~\bibnamefont
  {Dum}}, \bibinfo {author} {\bibfnamefont {P.}~\bibnamefont {Zoller}}, \ and\
  \bibinfo {author} {\bibfnamefont {H.}~\bibnamefont {Ritsch}},\ }\href@noop {}
  {\bibfield  {journal} {\bibinfo  {journal} {Physical Review A}\ }\textbf
  {\bibinfo {volume} {45}},\ \bibinfo {pages} {4879} (\bibinfo {year}
  {1992})}\BibitemShut {NoStop}%
\end{thebibliography}%
%\bibliography{main.bib}

\end{document}